\documentclass[11pt]{article}
\usepackage[T1]{fontenc}
\usepackage[utf8]{inputenc}
\usepackage[a4paper,margin=2.2cm]{geometry}
\usepackage{amsmath,amssymb}
\usepackage{graphicx}
\usepackage{float}
\usepackage{xurl}
\usepackage{microtype}
\usepackage{authblk}
\usepackage{hyperref}
\usepackage{array}
\usepackage{booktabs}
\usepackage{caption}
\usepackage{enumitem}
\hypersetup{colorlinks=true,linkcolor=blue,citecolor=blue,urlcolor=blue}
\graphicspath{{figures/}}
\title{Anisotropic mesoscale magnetic scattering across the spin-reorientation transition in TbMn$_6$Sn$_6$}
\author[1,2,$\dagger$]{Zhuang Xu}
\author[1,2,$\dagger$]{Fan Chen}
\author[1,2]{Hui Cheng}
\author[1,2]{Xinmin Wang}
\author[1,2]{Taisen Zuo}
\author[1,2]{He Cheng}
\author[1,2,4]{Haitao Hu}
\author[1,2]{Cuiping Zhang}
\author[1,2]{Qian Zhao}
\author[3]{Michel Kenzelmann}
\author[3,*]{Jonathan S. White}
\author[1,2,*]{Junying Shen}
\author[1,2,4,*]{Xin Tong}
\affil[1]{Spallation Neutron Source Science Center, Dongguan 523803, China}
\affil[2]{Institute of High Energy Physics, Chinese Academy of Sciences (CAS), Beijing 100049, China}
\affil[3]{Laboratory for Neutron Scattering and Imaging, Paul Scherrer Institute, CH-5232 Villigen, Switzerland}
\affil[4]{Guangdong Provincial Key Laboratory of Extreme Conditions, Dongguan 523803, China}
\affil[$\dagger$]{These authors contributed equally to this work.}
\affil[*]{Correspondence: jonathan.white@psi.ch (J.S.W.); shenjy@ihep.ac.cn (J.Y.S.); tongxin@ihep.ac.cn (X.T.)}
\date{}

\begin{document}
\maketitle
\begin{abstract}
TbMn$_6$Sn$_6$ is a centrosymmetric ferrimagnetic kagome metal that combines topological electronic states with strong magnetic anisotropy and a near-room-temperature spin-reorientation transition (SRT). Real-space imaging studies of thin specimens have revealed helical stripes, biskyrmion-like textures, and skyrmion-bubble-like structures in the spin-reorientation regime. Whether such local textures develop into a bulk long-range-ordered skyrmion lattice remains unresolved. Here we present small-angle neutron scattering (SANS) results probing the mesoscale magnetism of a bulk TbMn$_6$Sn$_6$ single crystal. We observe no sixfold SANS Bragg pattern indicative of a long-range-ordered skyrmion lattice under the explored conditions. Instead, the dominant magnetic response is an anisotropic twofold low-$Q$ scattering component whose intensity distribution and field dependence evolve markedly across the spin-reorientation regime. In zero field, a broad and horizontally elongated low-$Q$ signal develops above $T_{\rm SR}$ and weakens as the system approaches the paramagnetic state. At 345 K with $H\parallel b$, the signal is diffuse and is suppressed rapidly below about 700 Oe. By contrast, at 310 K with $H\parallel c$, close to $T_{\rm SR}$, the response is more concentrated and line-shaped and persists to higher fields. The contrasts between these scans likely reflect the combined effects of rapidly evolving magnetic anisotropy near $T_{\rm SR}$ and the field geometry. The angular intensity distribution for $H\parallel c$ is additionally influenced by the magnetic SANS orientation factor. The intensity-weighted characteristic scattering vector $q^*$ and the corresponding $d=2\pi/q^*$ are used as descriptive measures of the diffuse reciprocal-space distribution. Our results therefore establish a bulk anisotropic mesoscale magnetic response across the spin-reorientation regime, consistent with stripe- or domain-related textures reported by real-space imaging, while not uniquely identifying their real-space morphology.
\end{abstract}


\section{Introduction}
Topological magnets provide a platform for connecting real-space spin textures with momentum-space electronic topology. In kagome magnets, this interplay is particularly rich because geometrical frustration, spin-orbit coupling, exchange interactions and magnetic anisotropy can operate on comparable energy scales. TbMn$_6$Sn$_6$ is a prominent member of the $R$Mn$_6$Sn$_6$ family ($R$ = rare earth). It crystallizes in the centrosymmetric hexagonal $P6/mmm$ structure, consisting of Mn kagome layers separated by Tb-Sn layers. The material exhibits ferrimagnetic order below approximately 423 K, and its low-temperature out-of-plane magnetic state is associated with Chern-gapped kagome electronic bands and quantum-limit Chern magnetism \cite{ref1}. A defining feature of TbMn$_6$Sn$_6$ is that its magnetic anisotropy changes strongly with temperature. The easy magnetization axis switches from the $c$ axis at low temperature to the basal plane through a spin-reorientation process near room temperature, $T_{\rm SR}\sim310$ K \cite{ref2,ref3,ref4}.

Because the spin-reorientation transition reflects a changing balance among exchange, magnetocrystalline anisotropy, magnetostatic energy and Zeeman energy, it can stabilize mesoscale magnetic textures. Lorentz transmission electron microscopy (L-TEM) has directly observed periodic helical stripe domains in TbMn$_6$Sn$_6$ over a broad temperature range near the spin-reorientation transition and biskyrmion-like textures in a narrower temperature and magnetic-field range \cite{ref2,ref3}. In those works, the helix periodicity was reported to be about 100 nm at 310 K, and the texture evolved to a biskyrmion structure followed by a polarized state under external magnetic field. A subsequent L-TEM study further shows that skyrmion bubbles can be generated near the spin-reorientation domain interface by a convergent electron beam and that the generation is attributed to local anisotropy variations, dipolar effects and domain-wall confinement rather than to a conventional Dzyaloshinskii-Moriya interaction \cite{ref3}. Magneto-optic Kerr microscopy has also imaged stripy magnetic domains in TbMn$_6$Sn$_6$ and found that the domain contrast degrades with increasing temperature and disappears above the spin-flip temperature \cite{ref4}. These observations demonstrate that mesoscale domains are strongly coupled to the temperature-dependent magnetic anisotropy of TbMn$_6$Sn$_6$.

Nevertheless, real-space imaging on thin lamellae and bulk reciprocal-space measurements do not necessarily probe the same magnetic state. Since TbMn$_6$Sn$_6$ is centrosymmetric, bulk Dzyaloshinskii-Moriya interactions (DMI) are symmetry-forbidden. The skyrmion-like textures reported in thin specimens are instead associated with a balance among competing exchange interactions, including RKKY-type interactions, dipole--dipole interactions, anisotropy, Zeeman coupling and thickness-dependent magnetostatic energy \cite{ref2,ref3}. Such energy contributions are sensitive to sample thickness, demagnetizing fields, and boundary conditions. A bulk-sensitive method is therefore needed to determine whether the local magnetic textures observed in thin specimens develop into a bulk long-range-ordered skyrmion lattice or instead remain as finite-range stripe-like or domain-wall correlations in a bulk crystal.

Magnetic SANS is well suited for this problem because it probes bulk magnetic correlations through the Fourier components of the magnetization distribution on mesoscopic length scales, typically from a few to several hundred nanometers \cite{ref5,ref6,ref7}. In the small-angle regime, a one-dimensional helix or stripe-like modulation produces a pair of scattering maxima at approximately $\pm q$, whereas an ordered hexagonal skyrmion lattice produces sixfold Bragg peaks when the lattice plane is oriented perpendicular to the neutron beam \cite{ref5,ref8}. In more disordered situations, SANS may observe rings, arcs or broad twofold diffuse features rather than sharp spots. Simulations and SANS experiments on disordered skyrmion materials show that labyrinth domains, multidomain helices and rotationally disordered skyrmion states can produce broadened or ring-like reciprocal-space features \cite{ref9,ref10}. Similarly, SANS studies of ferrimagnetic spinels and magnetic multilayers have demonstrated that stripe-like domain-wall patterning can extend throughout a bulk or multilayer sample and can give broad anisotropic small-angle magnetic scattering rather than sharp diffraction peaks \cite{ref11,ref12}.

Several related systems provide useful context. In centrosymmetric MnNiGa, L-TEM reported biskyrmion nanodomains over a wide temperature range, whereas SANS measurements on bulk single crystals found no evidence for a long-range biskyrmion lattice; nevertheless, the internal twofold axes remained aligned by magnetocrystalline anisotropy \cite{ref13,ref14}. In FeGe thin films, the field response of the helical modulation was found to depend strongly on the field direction \cite{ref15}, while studies of Co$_8$Zn$_8$Mn$_4$ and (Fe,Co)Si showed that disorder, pinning and field history can reorganize mesoscale magnetic textures without producing a simple equilibrium lattice \cite{ref9,ref16}. These examples motivate the present bulk study without prescribing a unique real-space texture. We use SANS to determine how the anisotropic mesoscale magnetic response of TbMn$_6$Sn$_6$ evolves through the spin-reorientation regime and to test whether any long-range sixfold skyrmion-lattice diffraction develops in the bulk.

\section{Materials and Methods}
High-quality TbMn$_6$Sn$_6$ single crystals were grown by the self-tin-flux method \cite{ref1,ref2,ref3,ref4}. The bulk single crystal used for SANS had a mass of 121 mg and dimensions of approximately $6.7\,\mathrm{mm}\times5.7\,\mathrm{mm}\times0.6\,\mathrm{mm}$. The crystallographic orientation was checked by Laue diffraction, and the phase purity and orientation were verified by X-ray diffraction. Magnetization measurements were performed with magnetic field applied along representative crystallographic directions in order to locate the spin-reorientation transition and to select suitable temperature and field windows for SANS.

SANS experiments were performed on the SANS-I instrument of the Swiss Spallation Neutron Source, Paul Scherrer Institute. The incident neutron wavelength was close to 11~\AA{} with a wavelength spread of approximately 10\%, and the main sample-detector distance was about 18 m. The neutron beam was aligned along the crystallographic $b$ axis so that the detector plane corresponded approximately to the $ac$ reciprocal-space plane. This geometry and the temperature and magnetic-field ranges were chosen with reference to L-TEM studies that reported helical/stripe and biskyrmion-like textures near the spin-reorientation regime \cite{ref2,ref3}. We measured a zero-field temperature scan across a wider interval than the imaging studies and field scans for both $H\parallel b$ and $H\parallel c$, thereby characterizing the temperature and field evolution of the low-$Q$ scattering, its anisotropy, and field suppression while also searching for signatures of a bulk long-range-ordered skyrmion lattice beyond the previously reported texture window. For $H\parallel c$, the angular intensity on the detector also reflects the magnetic SANS orientation factor, because only magnetization components perpendicular to the scattering vector contribute to the magnetic scattering cross section \cite{ref5,ref6}.

The SANS data were reduced using GRASP \cite{ref17} and inspected using GRASP-based reciprocal-space visualization procedures \cite{ref18}. Detector efficiency correction, monitor normalization, masking and transformation to $Q_x$-$Q_y$ coordinates were carried out consistently for all scans. The data were intensity-normalized and background-subtracted before further analysis. For each region of anisotropic magnetic scattering, the integrated intensity $I$ used in the main figures was obtained by summing the background-subtracted detector intensity over the corresponding two-dimensional region of interest (ROI). Because the observed low-$Q$ magnetic signal is diffuse rather than resolution-limited Bragg peaks, the characteristic wave vector was extracted from the intensity-weighted first moment of the background-subtracted intensity distribution within each ROI \cite{ref19,ref20,ref21}. We denote this quantity simply as $q^*$:
\begin{equation}
q^*=\frac{\sum_i I_i q_i}{\sum_i I_i},
\end{equation}
where $q_i=(Q_{x,i}^2+Q_{y,i}^2)^{1/2}$ is the magnitude of the scattering vector, and $I_i$ is the background-subtracted intensity at detector pixel $i$ within the ROI. The values from the two regions were averaged, and the corresponding inverse-$q$ descriptor was calculated as
\begin{equation}
d=\frac{2\pi}{q^*}.
\end{equation}
We use $d$ as a convenient inverse-$q$ descriptor of the radial position of the diffuse magnetic scattering. The data sets, backgrounds and ROI definitions, together with the ROI-variation and same-temperature high-field-reference checks, are summarized in the Supplementary Material.

\section{Results and Discussion}
\begin{figure}[htbp]
\centering
\includegraphics[width=0.96\linewidth]{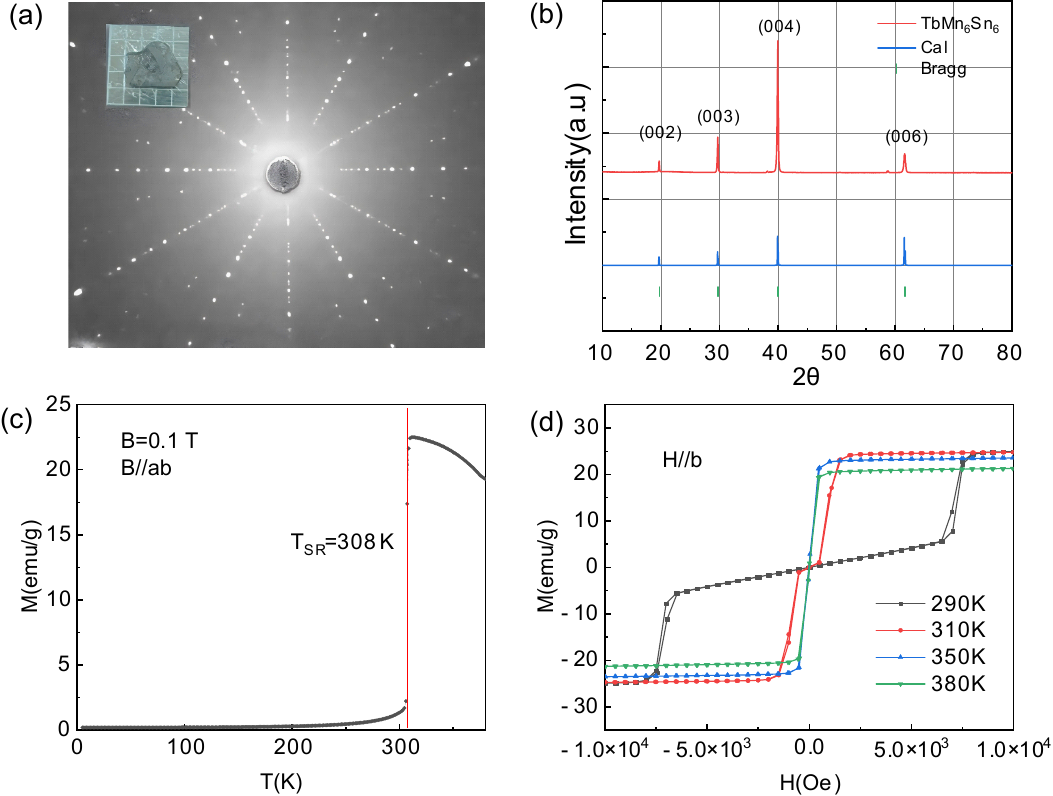}
\caption{Characterization of the TbMn$_6$Sn$_6$ single crystal. (a) Laue diffraction image and optical photograph of the crystal used for the SANS measurements. (b) Single-crystal X-ray diffraction pattern. (c) Temperature-dependent magnetization for $B\parallel ab$, identifying the spin-reorientation transition. (d) Field-dependent magnetization for $H\parallel b$ at representative temperatures.}
\label{fig:fig1}
\end{figure}

Figure~\ref{fig:fig1} summarizes the structural and magnetic characterization of the TbMn$_6$Sn$_6$ crystal. The Laue and X-ray diffraction data confirm the single-crystalline character and crystallographic orientation. The temperature-dependent magnetization identifies a spin-reorientation transition near 308 K, as reported in previous neutron, magnetization and imaging studies of TbMn$_6$Sn$_6$ \cite{ref1,ref2,ref3,ref4}. The magnetization curves with $H\parallel b$ in Figure~\ref{fig:fig1}(d) show that the field scale for in-plane polarization changes sharply through the spin-reorientation region. Above the transition the magnetization approaches the high-field regime at substantially lower fields than at 290 K, consistent with the strong temperature dependence of the magnetocrystalline anisotropy reported for TbMn$_6$Sn$_6$ \cite{ref22,ref23,ref24}.

\begin{figure}[htbp]
\centering
\includegraphics[width=0.98\linewidth]{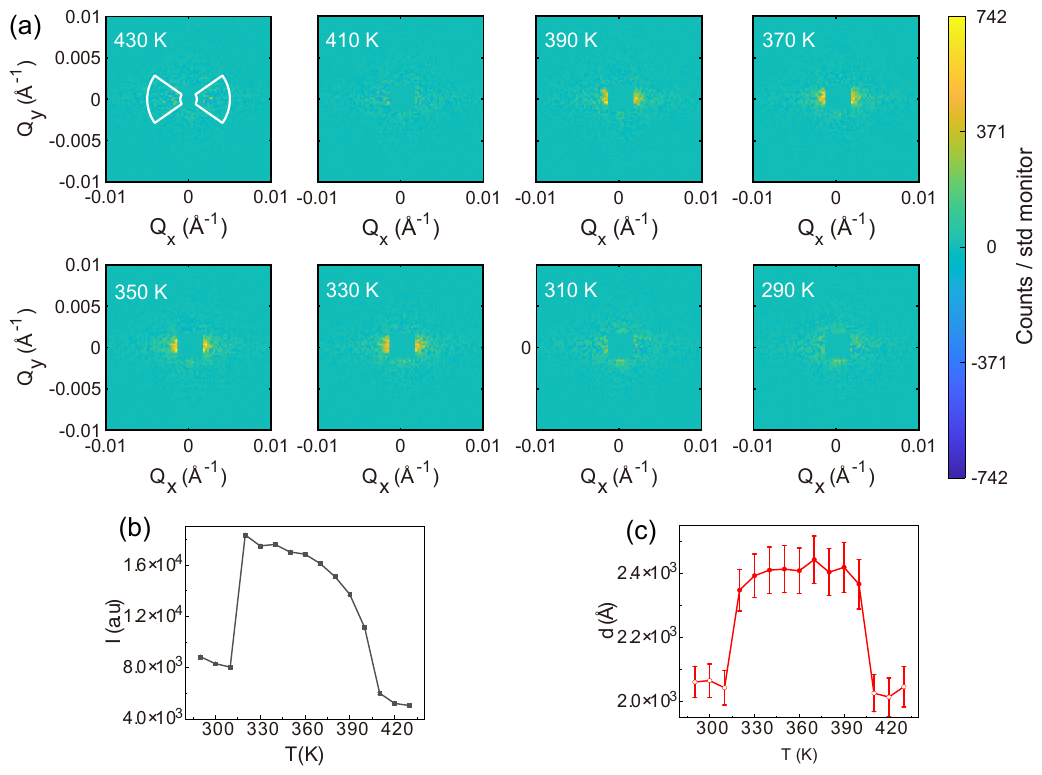}
\caption{Zero-field temperature scan in the beam $\parallel b$ geometry. (a) Background-subtracted SANS patterns. (b,c) Temperature dependences of $I$ and $d$. A broad horizontal twofold low-$Q$ signal appears above approximately 310 K and persists below about 400 K.}
\label{fig:fig3}
\end{figure}

We first consider the zero-field temperature scan with the beam along $b$. As shown in Figure~\ref{fig:fig3}(a), a broad horizontal twofold low-$Q$ signal emerges near the spin-reorientation temperature, is strongest between roughly 320 and 380 K, and fades on approaching 400 K. This is the same broad temperature interval in which L-TEM and MOKE studies reported helical or stripe-like domain textures \cite{ref2,ref3,ref4}. The coincidence in temperature and the twofold anisotropy make stripe- or domain-related correlations a natural real-space interpretation, but the SANS result itself is more generally described as anisotropic mesoscale magnetic scattering. The integrated intensity in Figure~\ref{fig:fig3}(b) follows the development and loss of this magnetic contrast. The inverse-$q$ descriptor $d$ in Figure~\ref{fig:fig3}(c) varies much less over the same interval. Its numerical values are comparable to the mesoscale spacings reported by real-space imaging \cite{ref2,ref3}; we therefore use it as a secondary descriptor of the diffuse scattering rather than as the defining evidence for a specific stripe periodicity. No sixfold pattern is observed under these conditions.

\begin{figure}[htbp]
\centering
\includegraphics[width=0.98\linewidth]{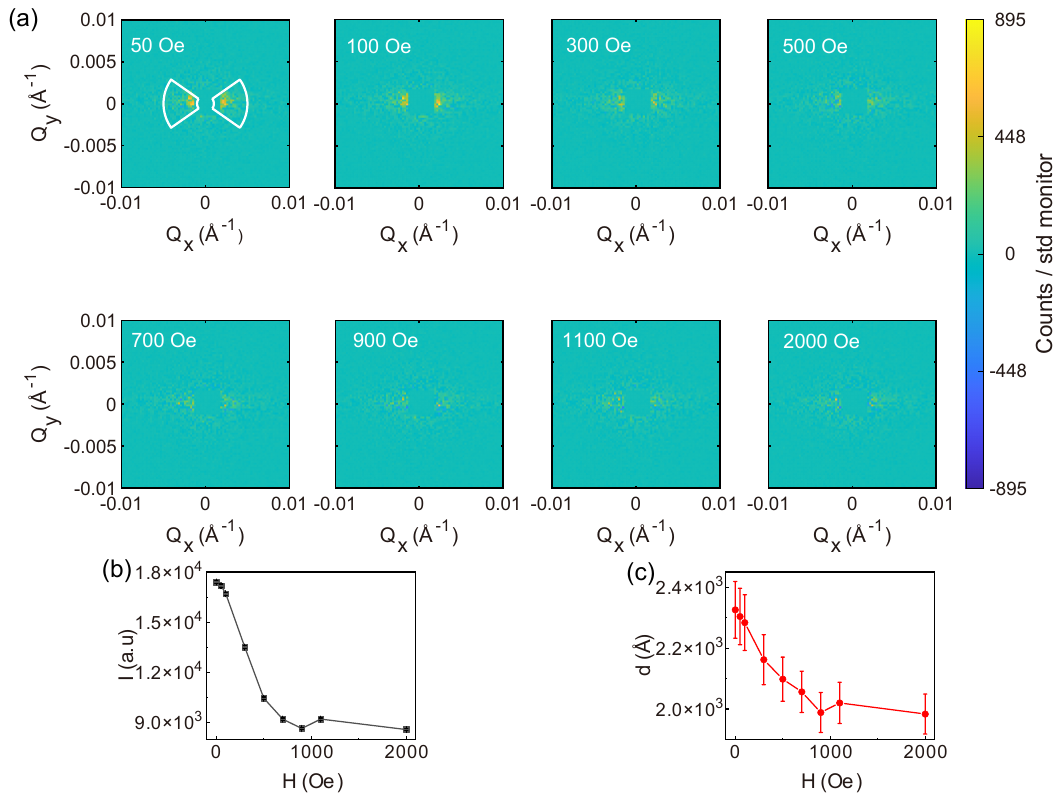}
\caption{Field scan at 345 K in the beam $\parallel b$, field $\parallel b$ geometry. (a) Background-subtracted SANS patterns. (b,c) Field dependences of $I$ and $d$. $I$ and $d$ decrease up to about 700 Oe and become nearly field independent from 700 to 2000 Oe.}
\label{fig:fig4}
\end{figure}

Figure~\ref{fig:fig4} shows the field scan at 345 K for beam $\parallel b$ and $H\parallel b$. The low-$Q$ response is broad and diffuse. Its integrated intensity falls rapidly up to about 700 Oe and changes only weakly at higher fields; $d$ decreases over the same low-field interval and then approaches a plateau. At 345 K, above $T_{\rm SR}$ and within the high-temperature easy-plane regime, the magnetic anisotropy has evolved toward the high-temperature easy-plane state \cite{ref22,ref23,ref24}. An in-plane field can therefore align the magnetization comparatively efficiently, providing a natural explanation for the rapid loss of the anisotropic low-$Q$ contrast. The evolution of $d$ is treated more cautiously. Rather than implying a literal contraction of an ideal stripe period, it is interpreted as being consistent with a redistribution of diffuse spectral weight as the larger-scale low-$q$ component is preferentially suppressed. The crossover field is comparable to the scale on which the $H\parallel b$ magnetization in Figure~\ref{fig:fig1}(d) approaches its high-field regime. This behaviour is broadly consistent with field-driven reorganization of mesoscale magnetic textures reported in other SANS studies \cite{ref11,ref12,ref15,ref16}.

\begin{figure}[htbp]
\centering
\includegraphics[width=0.98\linewidth]{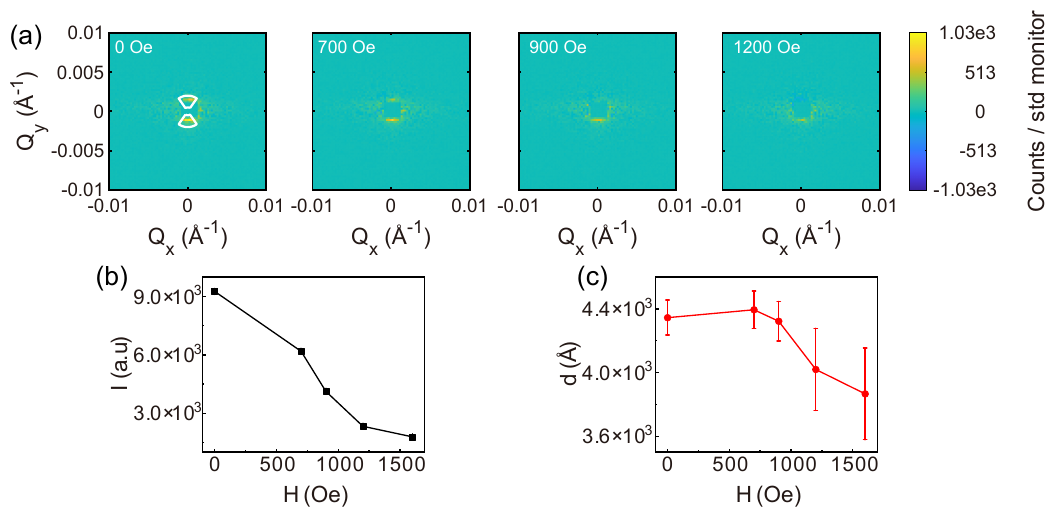}
\caption{Field scan at 310 K in the beam $\parallel b$, field $\parallel c$ geometry. (a) Background-subtracted SANS patterns. (b,c) Field dependences of $I$ and $d$. The low-$Q$ response is more concentrated and line-shaped than at 345 K and is suppressed more gradually over the measured field range.}
\label{fig:fig5}
\end{figure}

Figure~\ref{fig:fig5} presents the $H\parallel c$ scan at 310 K, close to $T_{\rm SR}$. The magnetic scattering is distinctly more concentrated and line-shaped than that observed at 345 K for $H\parallel b$, with its dominant intensity distributed approximately perpendicular to the $c$ axis. The signal also decays more gradually with increasing magnetic field and remains visible up to the highest measured field. Because the 310 K and 345 K measurements differ in both temperature and field direction, the contrast observed cannot be uniquely assigned to either variable alone. A physically consistent interpretation is that the rapidly evolving magnetic anisotropy near $T_{\rm SR}$ makes the mesoscale magnetization particularly susceptible to field orientation, while the $H\parallel c$ geometry weights a different set of magnetic Fourier components through the magnetic SANS orientation factor \cite{ref5,ref6,ref22,ref23,ref24}. The observed evolution from a broad horizontal response at 345 K to a concentrated, nearly orthogonal response at 310 K therefore indicates a strongly anisotropic reorganization of the mesoscale magnetic scattering, rather than providing direct evidence for a $90^\circ$ rotation of real-space stripes. The field dependence of $d$ is correspondingly used only as a descriptive measure of the evolving diffuse distribution. Its magnitude remains comparable to the mesoscale texture lengths reported in L-TEM studies \cite{ref2,ref3}.

\begin{figure}[htbp]
\centering
\includegraphics[width=0.98\linewidth]{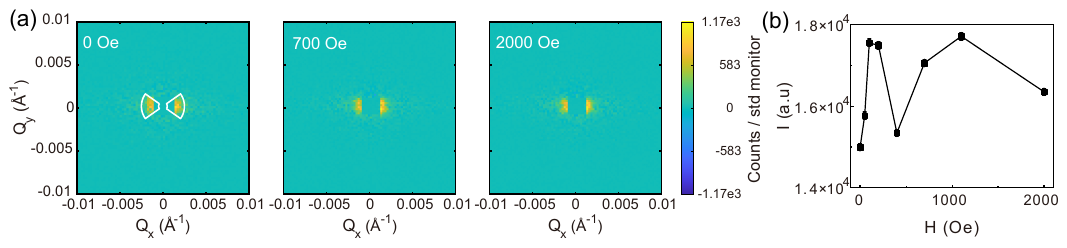}
\caption{Field scan at 400 K in the beam $\parallel b$, field $\parallel c$ geometry. (a) Representative background-subtracted SANS patterns. (b) Integrated intensity as a function of field. The response is less field dependent and returns toward a horizontal twofold character compared with the field scan along $c$ at 310 K.}
\label{fig:fig6}
\end{figure}

Figure~\ref{fig:fig6} shows the corresponding $H\parallel c$ scan at 400 K. The response is weaker, more diffuse and much less systematic in its field dependence than that at 310 K, with the dominant anisotropy returning toward the horizontal direction. This high-temperature behaviour provides an important control, as the line-shaped response observed for $H\parallel c$ at 310 K is not a generic consequence of the scattering geometry, but is more pronounced in the vicinity of the spin-reorientation regime.

Taken together, the temperature- and field-dependent measurements establish the main result of this work, namely, the bulk low-$Q$ magnetic scattering varies its intensity, angular distribution and field scale across the spin-reorientation regime. At 345 K for $H\parallel b$, the magnetic scattering response is broad and rapidly suppressed, whereas at 310 K for $H\parallel c$, the response is more concentrated, nearly orthogonal in reciprocal space and persists to higher fields. This contrast is consistent with the combined effects of the temperature-dependent magnetic anisotropy and field-direction selection. The coincidence with the temperature window of stripe/helical textures seen by real-space imaging supports a stripe- or domain-related microscopic picture, while SANS directly establishes the anisotropic mesoscale scattering rather than a unique real-space texture. No sixfold magnetic diffraction peaks are observed in any measured configuration.

The absence of sixfold diffraction peaks from a bulk long-range-ordered skyrmion lattice is compatible with the microscopic magnetic ingredients established for TbMn$_6$Sn$_6$. The centrosymmetric $P6/mmm$ structure forbids the uniform bulk DMI that provides the conventional route to long-wavelength chiral helimagnetism. The absence of conventional bulk DMI removes one common route for fixing chirality and stabilizing a long-range chiral skyrmion lattice, although frustrated exchange and dipolar interactions can support topological textures by other mechanisms. In TbMn$_6$Sn$_6$, nontrivial textures have instead been attributed to a balance among RKKY-type frustrated exchange, dipolar interaction, anisotropy variation, Zeeman energy and the spin-reorientation interface \cite{ref2,ref3}. These ingredients are sensitive to sample thickness, boundary conditions and lamella geometry. The electron-assisted study explicitly showed that homogeneous field and temperature changes do not always produce a bubble lattice, whereas a local SRT interface created by a convergent electron beam can generate skyrmion bubbles \cite{ref3}. This sensitivity is consistent with the present observation of anisotropic diffuse bulk scattering without a long-range ordered sixfold skyrmion-lattice pattern. More generally, skyrmion theory and experiments show that DMI, frustrated exchange and dipolar energies may all stabilize topological textures under suitable conditions, but positional and orientational order depend on anisotropy, dimensionality, disorder and pinning \cite{ref25,ref26,ref27,ref28,ref29,ref30,ref31,ref32}.

The diffuse nature of the twofold SANS signal indicates that the observed scattering is distributed over a finite region of reciprocal space rather than forming resolution-limited Bragg spots. The quantity $d$ therefore serves as a compact inverse-$q$ descriptor of the radial centroid of the diffuse scattering; its numerical similarity to spacings reported by L-TEM is suggestive, but does not by itself establish a specific real-space texture. The primary evidence in the present work is instead the anisotropic, temperature- and field-dependent magnetic scattering intensity. Broad anisotropic magnetic SANS has been widely observed in domain-patterned and spin-misaligned magnetic systems \cite{ref5,ref6,ref11,ref12}. This interpretation is consistent with classical stripe-domain and magnetic-domain theory, in which exchange, anisotropy, magnetostatic energy, Zeeman coupling and pinning jointly determine mesoscale domain patterns \cite{ref33,ref34,ref35}. Studies of TbMn$_6$Sn$_6$ and related $R$Mn$_6$Sn$_6$ compounds further reveal competing Mn--Mn and Mn--Tb exchange interactions, rare-earth single-ion anisotropy, and electronic correlations \cite{ref36,ref37,ref38,ref39}. Neutron diffraction, high-field magnetization and microscopic imaging studies have independently demonstrated the strong sensitivity of the spin-reorientation process to magnetic field and anisotropy \cite{ref40,ref41,ref42,ref43}. These established energy scales provide a natural framework for understanding the strongly anisotropic and field-dependent diffuse magnetic scattering observed here, without requiring a unique assignment to a particular real-space texture.

\section{Conclusions}
We have carried out a bulk-sensitive SANS study to characterize the mesoscale magnetic response of TbMn$_6$Sn$_6$ across its spin-reorientation regime. The dominant observation is a broad, anisotropic twofold low-$Q$ signal rather than a sixfold diffraction pattern from a long-range ordered skyrmion lattice. In zero field, the signal develops near $T_{\rm SR}$ and disappears on approaching the paramagnetic state. At 345 K with $H\parallel b$, it is diffuse and is suppressed rapidly below about 700 Oe. At 310 K with $H\parallel c$, it becomes more concentrated and line-shaped and is suppressed more gradually. Because these scans differ in both temperature and field orientation, their contrast is most naturally understood as the combined consequence of the rapidly evolving magnetic anisotropy near $T_{\rm SR}$ and the directional sensitivity of the applied-field geometry, including the magnetic SANS orientation factor. The quantity $d$ is used as a secondary inverse-$q$ descriptor of the diffuse reciprocal-space distribution rather than as a unique real-space stripe period. Its magnitude is comparable to the mesoscale spacings reported by L-TEM and MOKE, while the primary SANS result is the anisotropic magnetic scattering itself.

These results connect the spin-reorientation transition to a qualitative reorganization of magnetism on mesoscopic length scales. They also separate two levels of information that are often conflated: real-space imaging identifies specific textures in thin specimens, whereas SANS tests how such magnetism is expressed statistically throughout the illuminated bulk volume. Within the explored temperature, field and geometry range, TbMn$_6$Sn$_6$ shows no evidence for a coherent three-dimensional skyrmion lattice, but exhibits a pronounced anisotropic mesoscale response controlled by the spin-reorientation landscape.

\section*{Author contributions}
Conceptualization, J.Y.S.; methodology, Z.X., F.C., H.C., T.Z., H.C., H.H., J.S.W. and J.Y.S.; formal analysis, Z.X. and F.C.; investigation, Z.X., F.C., T.Z., X.W., C.Z. and Q.Z.; resources, J.Y.S., X.T. and J.S.W.; writing---original draft preparation, Z.X.; writing---review and editing, all authors; supervision, J.Y.S. and X.T.; funding acquisition, J.Y.S., X.T. and Z.X.

\section*{Data availability}
The data supporting the findings of this study are available from the corresponding authors upon reasonable request, subject to the data policies of the neutron facilities.

\section*{Acknowledgments}
This work was funded by the National Natural Science Foundation of China (No. 12675398 and No. 12104254), the Joint Funds of the National Natural Science Foundation of China (Grant No. U2430209), the National Science Fund for Distinguished Young Scholars (No. 12425512), the Postdoctoral Special Funding (DG251I4A19), the Guangdong Provincial Key Laboratory of Extreme Conditions (2023B1212010002). We would like to acknowledge the beam time allocated on SANS-I at the Swiss spallation neutron source SINQ, Paul Scherrer Institute, Villigen, Switzerland, and the beam time on the Very Small Angle Neutron Scattering at the China Spallation Neutron Source (CSNS) (https://cstr.cn/31113.02.CSNS.VSANS ).

\section*{Conflict of interest}
The authors declare no conflicts of interest.

\clearpage
\setcounter{figure}{0}
\setcounter{table}{0}
\renewcommand{\thefigure}{SM\arabic{figure}}
\renewcommand{\thetable}{S\arabic{table}}
\section*{Supplementary Material}

Supplementary data-set definitions, ROI-robustness checks and same-temperature high-field-reference analyses supporting the small-angle neutron scattering study of TbMn$_6$Sn$_6$ are provided. The analysis focuses on two-dimensional robustness checks and descriptive quantities used in the main text.
\section*{Data sets, backgrounds and analysis regions}
Table~\ref{tab:S1} lists the temperature/field ranges, background/reference conditions and ROI sectors used for the main-text figures and supplementary robustness checks. The sector notation is $[q_{\min},q_{\max},\theta,\Delta\theta]$, where $q_{\min}$ and $q_{\max}$ are the radial limits in \AA$^{-1}$, $\theta$ is the sector centre angle in the detector $Q_x$--$Q_y$ plane, and $\Delta\theta$ is the angular sector width in degrees.

\begin{table}[H]
\centering
\scriptsize
\caption{Summary of data sets, background/reference patterns and ROI sectors used in the SANS analysis.}
\label{tab:S1}
\setlength{\tabcolsep}{2pt}
\begin{tabular}{p{0.12\linewidth}p{0.22\linewidth}p{0.18\linewidth}p{0.22\linewidth}p{0.20\linewidth}}
\hline
Figure & Geometry & Temperature/field range & Background & ROI sector(s) \\
\hline
Main Fig. 2 & 0 Oe temperature scan, beam $\parallel b$ & 290--430 K, 0 Oe & 440 K, 0 Oe & $[0.001,0.005,0,70]$ and opposite sector \\
Main Fig. 3 / SM2 & 345 K, $H\parallel b$, beam $\parallel b$ & 0--2000 Oe & 455 K, 0 Oe; SM2 uses 2000 Oe at 345 K as the high-field reference & $[0.001,0.005,0,70]$ and opposite sector \\
Main Fig. 4 / SM3 & 310 K, $H\parallel c$, beam $\parallel b$ & 0--1600 Oe & 455 K, 0 Oe; SM3 uses 1600 Oe at 310 K as the high-field reference & $[0.0005,0.002,90,70]$ and opposite sector \\
Main Fig. 5 / SM4 & 400 K, $H\parallel c$, beam $\parallel b$ & 0--2000 Oe & 455 K, 0 Oe; SM4 uses 2000 Oe at 400 K as the high-field reference & $[0.0005,0.0025,0,70]$ and opposite sector \\
SM1(a--c) & ROI rotation test, 345 K, $H\parallel b$ & 0--2000 Oe & 455 K, 0 Oe & $[0.001,0.005,10,70]$ and opposite sector \\
SM1(d--f) & ROI angular-width test, 345 K, $H\parallel b$ & 0--2000 Oe & 455 K, 0 Oe & $[0.001,0.005,10,50]$ and opposite sector \\
SM1(g--i) & ROI radial-range test, 345 K, $H\parallel b$ & 0--2000 Oe & 455 K, 0 Oe & $[0.001,0.006,0,70]$ and opposite sector \\
\hline
\end{tabular}
\end{table}

The main text retains the integrated anisotropic intensity as the primary observable and $d=2\pi/q^*$ as a secondary inverse-$q$ descriptor derived from the two-dimensional intensity-weighted radial first moment. Because the low-$Q$ distribution is truncated by the beamstop, the absolute $d$ values are not used to establish an experimentally resolved finite-$q$ peak or a real-space stripe periodicity. The Supplementary checks are used to assess whether the qualitative trends survive reasonable changes of ROI and reference condition.

\section*{ROI and high-field-difference robustness checks}
\begin{figure}[H]
\centering
\includegraphics[width=0.98\linewidth]{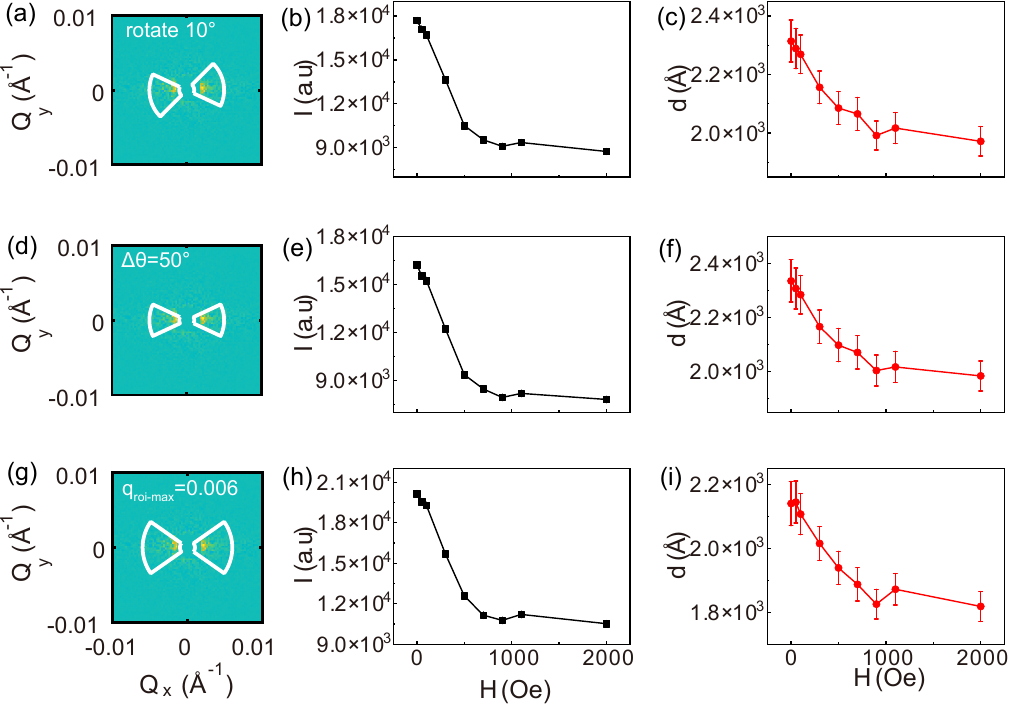}
\caption{ROI-sector robustness check for the 345 K, $H\parallel b$, beam $\parallel b$ field scan, using the 455 K, 0 Oe pattern as background. (a--c) Sector rotated by $10^\circ$, $[0.001,0.005,10,70]$. (d--f) Smaller angular width, $[0.001,0.005,10,50]$. (g--i) Larger radial range, $[0.001,0.006,0,70]$. The intensity trend is robust; the displayed $d=2\pi/q^*$ values are descriptive analysis quantities rather than a directly measured stripe period.}
\label{fig:SM1}
\end{figure}

Figure~\ref{fig:SM1} tests the sensitivity of the 345~K, $H\parallel b$, beam $\parallel b$ field scan to the two-dimensional ROI definition. Panels SM1(a)--SM1(c) use a sector rotated by $10^\circ$, panels SM1(d)--SM1(f) use a narrower angular width, and panels SM1(g)--SM1(i) use an expanded radial range. The integrated-intensity field dependence remains qualitatively unchanged under these reasonable ROI variations. The accompanying $q^*$-derived $d$ values are shown only as descriptive robustness diagnostics; their absolute values are more sensitive to the radial ROI range and are not interpreted as a stripe period.

\begin{figure}[H]
\centering
\includegraphics[width=0.98\linewidth]{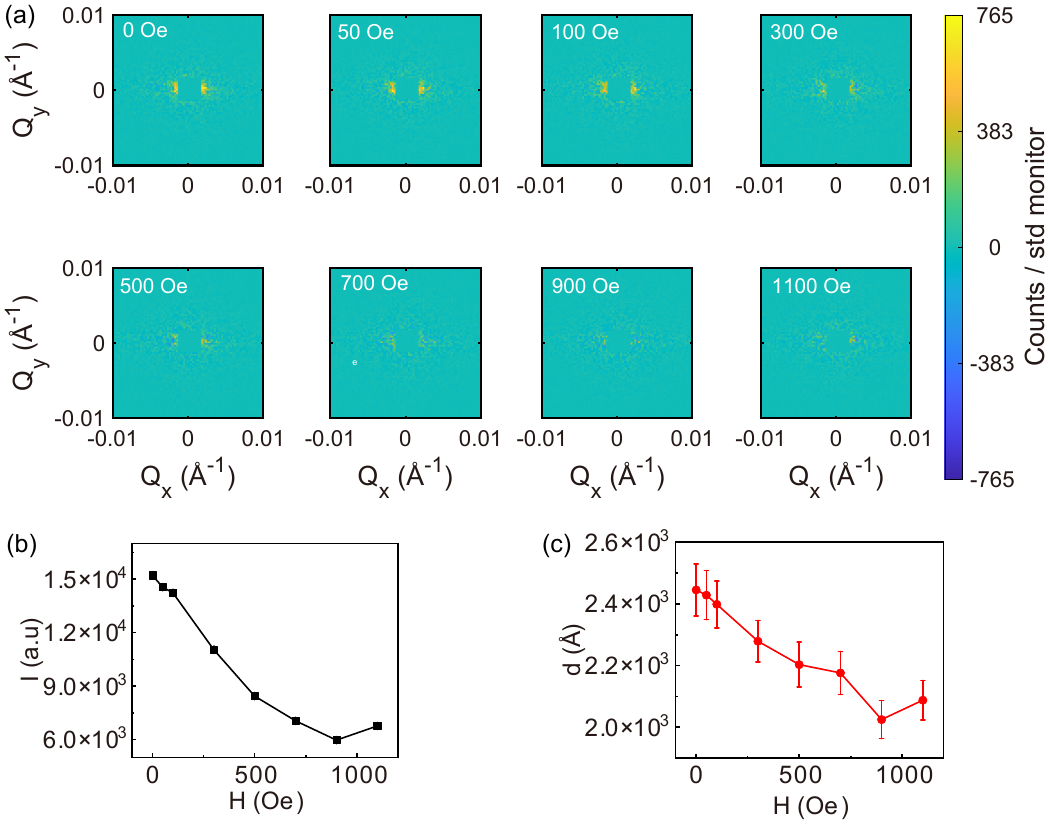}
\caption{High-field-reference analysis for the 345~K, $H\parallel b$, beam $\parallel b$ field scan. The 2000~Oe pattern at 345~K is used as the reference instead of the high-temperature background. (a) Difference SANS patterns relative to the 2000~Oe reference. (b) Integrated anisotropic intensity. (c) Descriptive $d=2\pi/q^*$ derived from the two-dimensional intensity-weighted radial first moment; it is not interpreted as a stripe periodicity.}
\label{fig:SM2}
\end{figure}

\begin{figure}[H]
\centering
\includegraphics[width=0.98\linewidth]{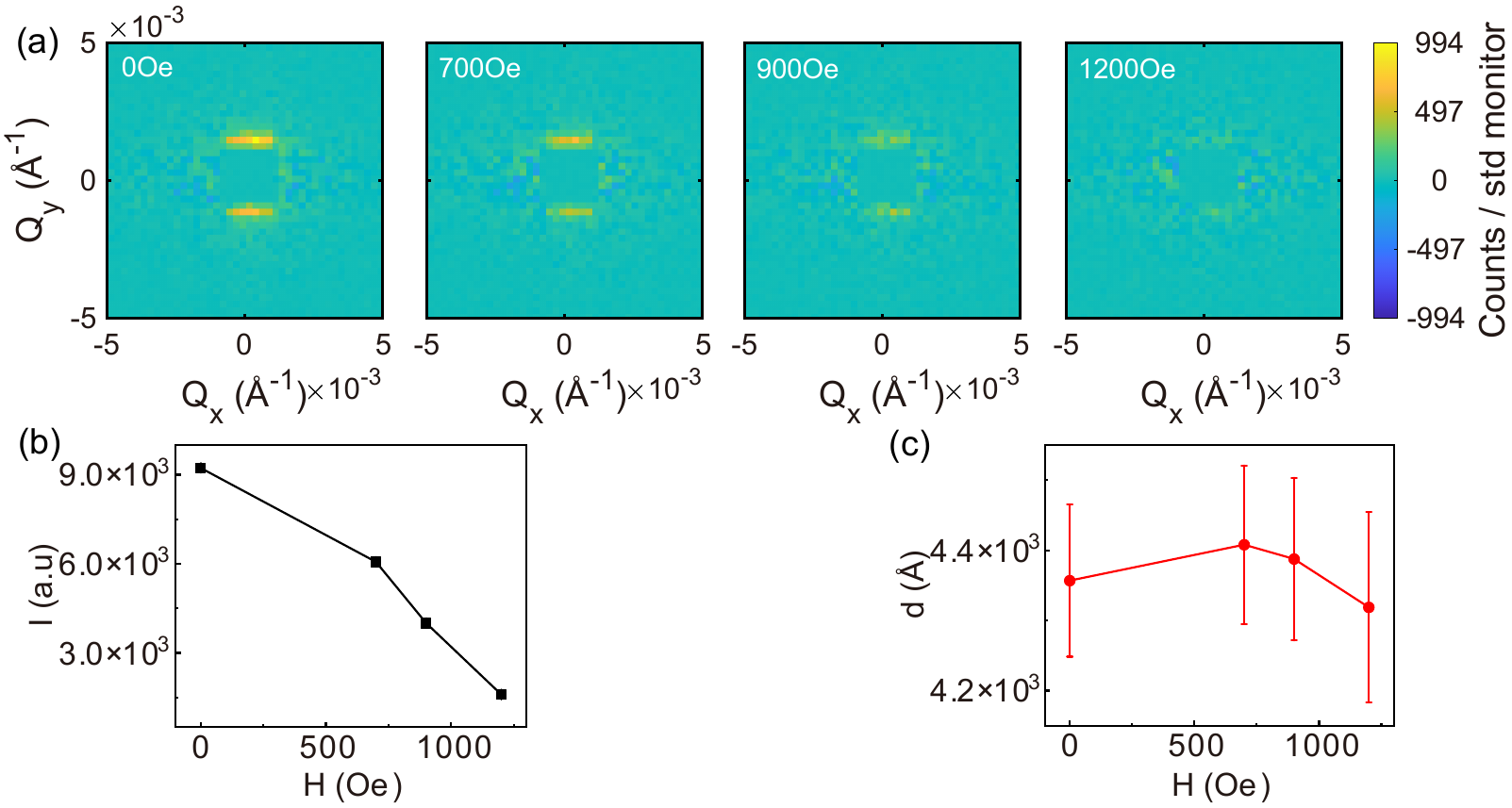}
\caption{High-field-reference analysis for the 310~K, $H\parallel c$, beam $\parallel b$ field scan. The 1600~Oe pattern at 310~K is used as the reference. (a) Difference SANS patterns relative to the 1600~Oe reference. (b) Integrated anisotropic intensity. (c) Descriptive $d=2\pi/q^*$ from the two-dimensional radial first moment.}
\label{fig:SM3}
\end{figure}

\begin{figure}[H]
\centering
\includegraphics[width=0.98\linewidth]{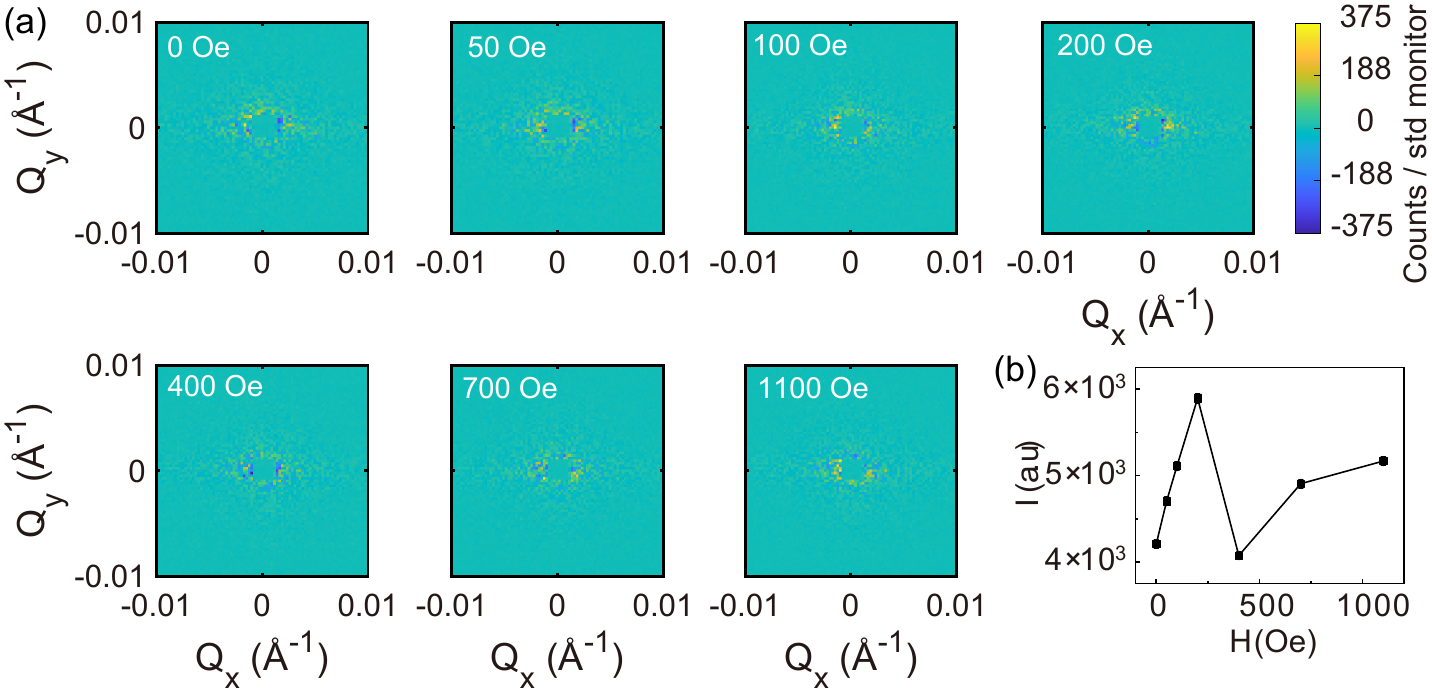}
\caption{High-field-reference analysis for the 400~K, $H\parallel c$, beam $\parallel b$ field scan. The 2000~Oe pattern at 400~K is subtracted from the lower-field patterns. (a) Difference SANS patterns relative to the 2000~Oe reference. (b) Integrated residual intensity. The irregular field dependence is consistent with a background-dominated residual, and no characteristic $q^*$ or $d$ is assigned.}
\label{fig:SM4}
\end{figure}

Figures~\ref{fig:SM2}--\ref{fig:SM4} provide same-temperature high-field-reference checks for the main field-dependent SANS analyses. The monitor normalization, beamstop masking and ROI definitions are the same as those used for the corresponding main-text data. Only the reference pattern is changed: a high-field pattern measured at the same temperature is subtracted from the lower-field patterns. Such a subtraction is a standard way to emphasize the field-dependent magnetic contribution when the reference approaches a high-field state, although the finite reference field need not represent perfect saturation \cite{smref1}. Because the reference subtraction changes the intensity baseline and the relative spectral weight within the ROI, the absolute values of descriptive centroid quantities need not match those obtained with the high-temperature background. The primary robustness criterion is the persistence of the anisotropic field-dependent intensity.

For the 345~K, $H\parallel b$, beam $\parallel b$ scan in Fig.~\ref{fig:SM2}, the 2000~Oe pattern is used as the high-field reference. The low-field difference patterns retain a horizontal anisotropic low-$Q$ component that is progressively suppressed with increasing field, consistent with the corresponding main-text intensity trend. This agreement supports the conclusion that the field evolution in main-text Fig.~3 is not generated by the choice of high-temperature background. The accompanying $d=2\pi/q^*$ values are retained only for descriptive comparison and are not used to infer a field-dependent stripe period.

For the 310~K, $H\parallel c$, beam $\parallel b$ scan in Fig.~\ref{fig:SM3}, the 1600~Oe pattern is used as the same-temperature reference. The difference patterns retain a field-dependent anisotropic low-$Q$ component whose intensity is reduced as the field approaches the reference state, supporting the main-text conclusion that the 310 K response is magnetic and field dependent. Since the 1600~Oe state may still contain magnetic scattering, this subtraction is interpreted only as a relative field-dependent contrast and not as an absolute saturated-state magnetic cross section. The displayed $d$ values are secondary inverse-$q$ descriptors and are not used to establish a real-space stripe period.

For the 400~K, $H\parallel c$, beam $\parallel b$ scan in Fig.~\ref{fig:SM4}, the 2000~Oe pattern is used as the same-temperature reference. After subtraction, the remaining intensity is weak and background-like. In contrast to Figs.~\ref{fig:SM2} and \ref{fig:SM3}, the integrated residual intensity varies irregularly with field rather than showing a systematic magnetic suppression. No stable anisotropic magnetic lobe can therefore be defined for this high-temperature difference data set, and no corresponding inverse-$q$ descriptor $d$ is reported.

{\renewcommand{\refname}{Supplementary References}
\makeatletter
\renewcommand\@biblabel[1]{[S#1]}
\makeatother

}

\end{document}